\documentstyle[epsfig, psfrag, 12pt, a4]{article}
\slshape
\newcommand{\beq}{\begin{equation}}
\newcommand{\eeq}{\end{equation}}
\newcommand{\be}{\begin{displaymath}}
\newcommand{\ee}{\end{displaymath}}
\newcommand{\bes}{\begin{eqnarray}}
\newcommand{\ees}{\end{eqnarray}}
\newcommand{\bee}{\begin{eqnarray*}}
\newcommand{\eee}{\end{eqnarray*}}

\newcommand{\margen}{\hspace{8mm}}

\begin{document}
\rightline{hep-th/9711068}
\vspace{8mm}
\centerline{\LARGE \bf THE BLACK HOLE:}
\vspace{3mm}
\centerline{\LARGE \bf SCATTERER, ABSORBER AND}
\vspace{2mm}
\centerline{\LARGE \bf EMITTER OF PARTICLES}
\vspace{4mm}
\begin{center} {{\bf N. SANCHEZ} \\  \vspace{2mm} \small {\it Observatoire de Paris-DEMIRM,} \\ \small  {\it 61 Avenue de l'Observatoire, 75014 PARIS, FRANCE}}
\end{center}
\begin{abstract}
Accurate and powerful computational methods developped by the author, based 
on the analytic resolution 
of the wave equation in the black hole background, allow to obtain the 
highly non trivial {\bf total absorption spectrum} of the Black Hole. 
As well as phase shifts and cross sections (elastic and inelastic) for a wide
range of energy and angular momentum, the angular distribution of absorbed 
and scattered waves, and the Hawking emission rates.
The total absorption spectrum of waves by the Black Hole is known exactly.
It presents as a function of frequency a remarkable
{\bf {oscillatory}} behaviour characteristic of a diffraction pattern.
It oscillates around its optical geometric limit (${{27}\over{4}} \pi {r_s}^2$)
with decreasing amplitude and almost constant period. This is an 
{\bf{unique}}
distinctive feature of the black hole absorption, and due to its $r=0$ singularity.
Ordinary absorptive bodies and optical models do not present these features.

The Hamiltonian describing the wave-black hole interaction is non hermitian
(despite being real) due to its singularity at the origin ($r=0$). The
unitarity optical theorem of scattering theory is generalized to the 
black hole case explicitely showing that absorption takes place only at 
the origin ($r = 0 $). 

All these results allow to {\bf{understand}}
and {\bf{reproduce}} the Black Hole absorption spectrum in terms of
Fresnel-Kirchoff diffraction theory: interference takes place between the
absorbed rays arriving at the origin by different optical paths.

These fundamental features of the Black Hole Absorption will be present
for generic higher dimensional Black Hole backgrounds, and whatever the low
energy effective theory they arise from.

In recent and increasing litterature devoted to compute absorption cross 
sections (``grey body factors'') of black holes (whatever ordinary, stringy,
D-braned), the fundamental remarkable features of the Black Hole 
Absorption spectrum are overlooked.
 \end{abstract}

\pagebreak[4]

\tableofcontents
%\listoffigures
\pagebreak[4]

\centerline{\LARGE \bf THE BLACK HOLE:}
\vspace{3mm}
\centerline{\LARGE \bf SCATTERER, ABSORBER AND}
\vspace{2mm}
\centerline{\LARGE \bf EMITTER OF PARTICLES}
\vspace{8mm}
\centerline{ \bf N. SANCHEZ}
\vspace{2mm}
\centerline{\it Observatoire de Paris-DEMIRM}
\centerline{\it  61, Avenue de l'Observatoire, 75014 Paris - FRANCE}
%\vspace{-2mm}

\section{INTRODUCTION AND RESULTS}
\margen I shall report here about some of my results on the physics of black holes and the dynamics of fields in the vicinity of such objects, describing at the same time, the Black Hole under its triple aspect of Scatterer, Absorber and Emitter
 of particles.

 I shall first report about the Absorption, it appears in the concept of black hole itself, the gravitational field being so intense that even light can not escape of it. Absorption is one of the properties that characterizes the black hole description in classical physics: black holes absorb waves but they can not emit them. If a quantum description of perturbation fields is considered, black holes also emit particles. For a static black hole, the quantum particle emission rate $H(k)$, and the classical wave absorption cross section $\sigma_A(k)$ are related by the Hawking's formula (1975, \cite{haw})
\beq
H(k) = {{\sigma_A(k)}\over{e^{8 \pi k M} - 1}} \; \; ,
\eeq
the factor relating them being planckien. Here $k$ and $M$ stands for the frequency of waves and for the mass of the  hole respectively.

We see the role played by the absorption in the emission. In spite of the extensive litterature discussing about the interaction of waves with black holes, the absorption spectrum $\sigma_A(k)$ (and other scattering parameters), as well as their theoretical foundations, was largely unsolved. In fact, the complexity of analytic solutions of the perturbation fields equations made this problem very difficult.

We have studied in detail the absorption spectrum of the black hole and we have found the total absorption cross section $\sigma_A(k)$ in the Hawking formula, obtaining a very simple expression (N. S\'{a}nchez, 1978 \cite{ns}), which is valid to very high accuracy over the entire range of k, namely

\bes \label{sigma}
\sigma_A(k) = 27 \pi M^2 - 2\sqrt{2} M {{\sin{(2\sqrt{27} \pi k M)}\over{k}}} & , \hspace{0.5cm} & kM \: \geq 0.07 \\
\sigma_A(0) =  16 \pi M^2  {\mathrm{\hspace{4.7cm}}} & &\nonumber
\ees

The absorption spectrum presents, as a function of the frequency, a remarkable oscillatory behaviour characteristic of a diffraction pattern (Fig.2). It oscillates around its constant geometrical optics value $ \sigma(\infty) = 27 \pi M^2$ with decreasing amplitude (as $ {1\over{(\sqrt{2} k M)}}$) and constant period ($ {2\over{3}} \sqrt{3} $). The value of $\sigma_A(0)$ is exactly given by $16 \pi M^2$. See below.

We have also calculated the Hawking radiation. This is only important in the interval $ 0 \leq k \leq {1\over{M}} $. The emission spectrum (Fig.1) does not show any of the interference oscillations characteristic of the absorption cross section, because the contribution of the S-wave dominates the Hawking radiation. The rapidly decrease of the Planck factor for $ kM \geq 1 $ supresses the contribution of higher partial waves.

Thus, for a black hole the emission follows a planckian spectrum, given by eq. (1), (Fig.1), and the absorption follows an oscillatory spectrum, given by eq.(2), (Fig.2).

\begin{figure}[h]
%\vspace{10cm}
\centering
\psfrag{k}{\bf {k}}
\psfrag{h}[r]{\bf{\it {H(k)}}$\ $}
\includegraphics[width=120mm, trim=0 170 0 150]{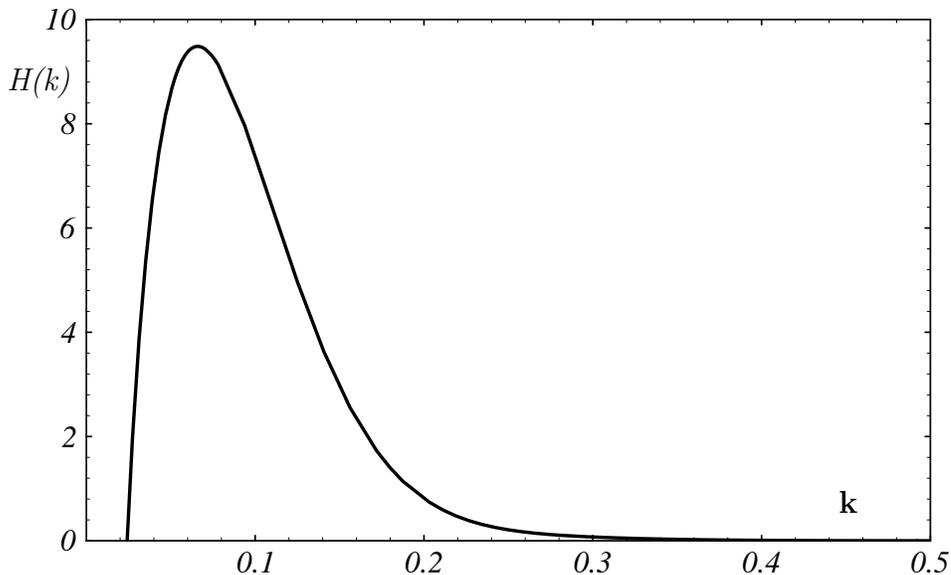}
\caption[fig1]{EMISSION BY A BLACK HOLE} \label{fig1}
\end{figure}

\begin{figure}[h]
%\vspace{10cm}
\centering
\psfrag{eqi}{${\sigma(\infty)=27 \pi M^2}$}
\psfrag{eq}{${\sigma_{A}(k)}$}
\psfrag{k}{\bf {k}}
\psfrag{s}[br]{${\sigma(\infty)}  \ $}
\includegraphics[width=120mm, trim=0 170 0 150]{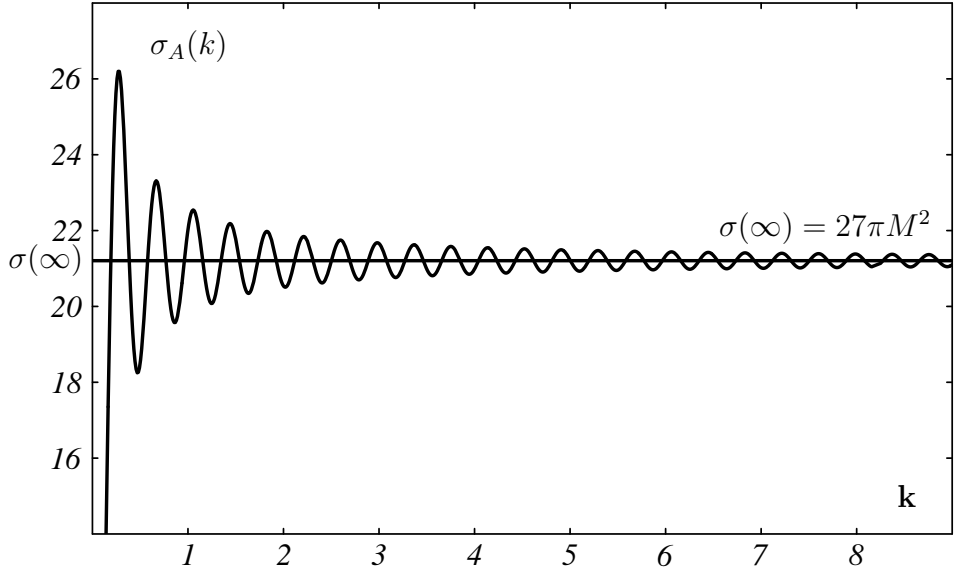}
\caption[Fig.II]{ABSORPTION BY A BLACK HOLE} \label{fig2}
\end{figure}

\begin{figure}[h]
%\vspace{6cm}
\centering
\psfrag{x}{$\:$ }
\psfrag{y}{$\:$ }
\psfrag{k}{\bf {k}}
\psfrag{s}{\bf ${\sigma_{A}(k)} \ $}
\psfrag{eq}{\bf ${\sigma(\infty)=\pi R^2}$}
\includegraphics[width=120mm, trim=0 170 0 150]{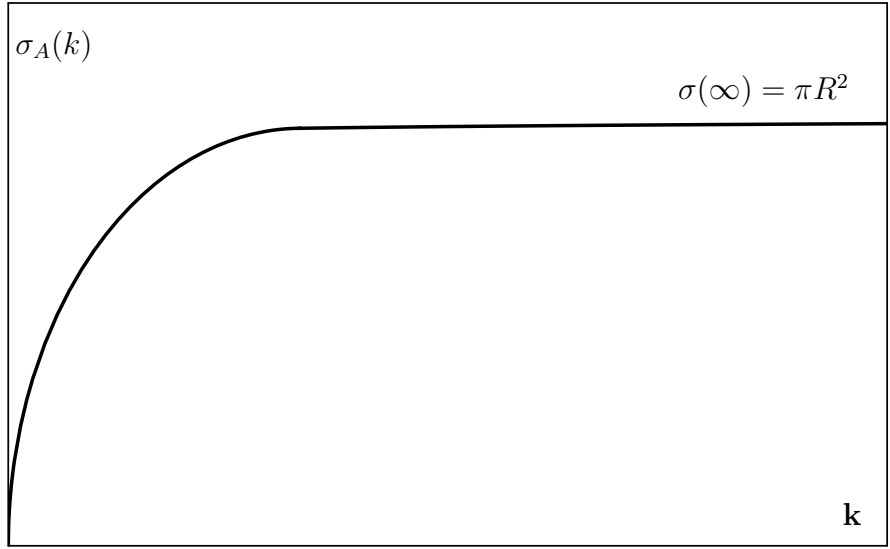}
\caption[Fig.III]{ABSORPTION BY A MATERIAL SPHERE WITH A COMPLEX REFRACTION INDEX} \label{fig3}
\end{figure}

It is interesting to compare the absorption by a black hole with that of other physical systems. Fig.3 shows the total absorption cross section for an ordinary material sphere with a complex refraction index. It is a monotonically increasing function of the frequency. It attains its geometrical optics limit without any oscillation. 

Comparison of Fig.2 with Fig.3 shows the differences between the absorptive properties of a black hole and those corresponding to ordinary absorptive bodies or described by complex potentials (optical models). For a black hole, the presence of absorption processes is due to the non hermitian character of the effective potential that describes the wave - black hole interaction \cite{ns77}. The effective Hamiltonian is non hermitian, despite of being real, due to its singularity at the origin ($ r = 0 $), as we have shown in ref.\cite{ns77}, so that the absorption takes place only at the origin. 

We have generalized to the black hole case the well known unitarity theorem (optical theorem) of the elastic potential scattering theory, explicitly relating the presence of a non zero absorption cross section to the existence of a singularity in the space-time \cite{ns77}. All these results allowed me to give a simple physical interpretation of the total absorption cross section $\sigma_A(k)$ in the context of the Fresnel-Kirchoff diffraction theory. The oscillatory behaviour of $\sigma_A(k)$ is explained in a good approximation by the interference of the absorbed rays arriving at the origin through different optical paths. \cite{ns}

Usually, in scattering theory, absorption processes are related to complex (and non-singular) potentials. On the contrary, in the black hole case, the potential is real and singular at the origin. All partial absorption amplitudes have absolute maxima at the frequence $k = {3\over4} ({{\sqrt{3}}\over{M}})(l + {1\over{2}})$. By summing up for all angular momenta, each absolute maximum of partial absorption cross section produces a relative maximum in the total absorption spectrum giving rise to the presence of oscillations.

It can be pointed out that associated to the planckian spectrum, the black hole has a temperature equal to ${1\over{(8 \pi M)}}$. In what concerns the absorption spectrum it is not possible to associate a refraction index to the black hole. For optical materials, the absorption takes place in the whole volume, whereas for the black hole, it takes place only at the origin.

It is also interesting to calculate the angular distribution of absorbed waves. For it one must study too the black hole as elastic scatterer.

The distribution of scattered waves, as a function of the scattering angle $\theta$, has been computed in a wide range of the frequency \cite{ns77},\cite{ns2}. It presents a strong peak $(\sim {\theta^{-4}})$ characteristic of long range interactions in the forward direction, and a ``glory'' in the backward, characteristic of the presence of strongly attractive interactions for short distances. For intermediate $\theta$, it shows a complicated behaviour with peaks and drops that disappear only at the geometrical-optics limit.

\begin{figure}[h!]
%\vspace{15cm}
\centering
\psfrag{t}[b]{\bf $\theta$}
\psfrag{g}[r]{${\mid g(\theta) \mid}^2 \ \ \ \ $}
%\psfrag{1}[b]{\it 1}
%\psfrag{2}[b]{\it 2}
%\psfrag{3}[b]{\it 3}
\psfrag{a}{$x_s = 0.1$}
\psfrag{b}{$x_s = 0.25$}
\psfrag{c}{$x_s = 0.5$}
\psfrag{d}[]{$\ \ \ \ \ \ \ \ \ x_s = 1.0$}
\psfrag{e}[]{$\ \ \ \ \ \ \  x_s = 2.0$}
\psfrag{i}{$\ $}
\includegraphics[width=145mm]{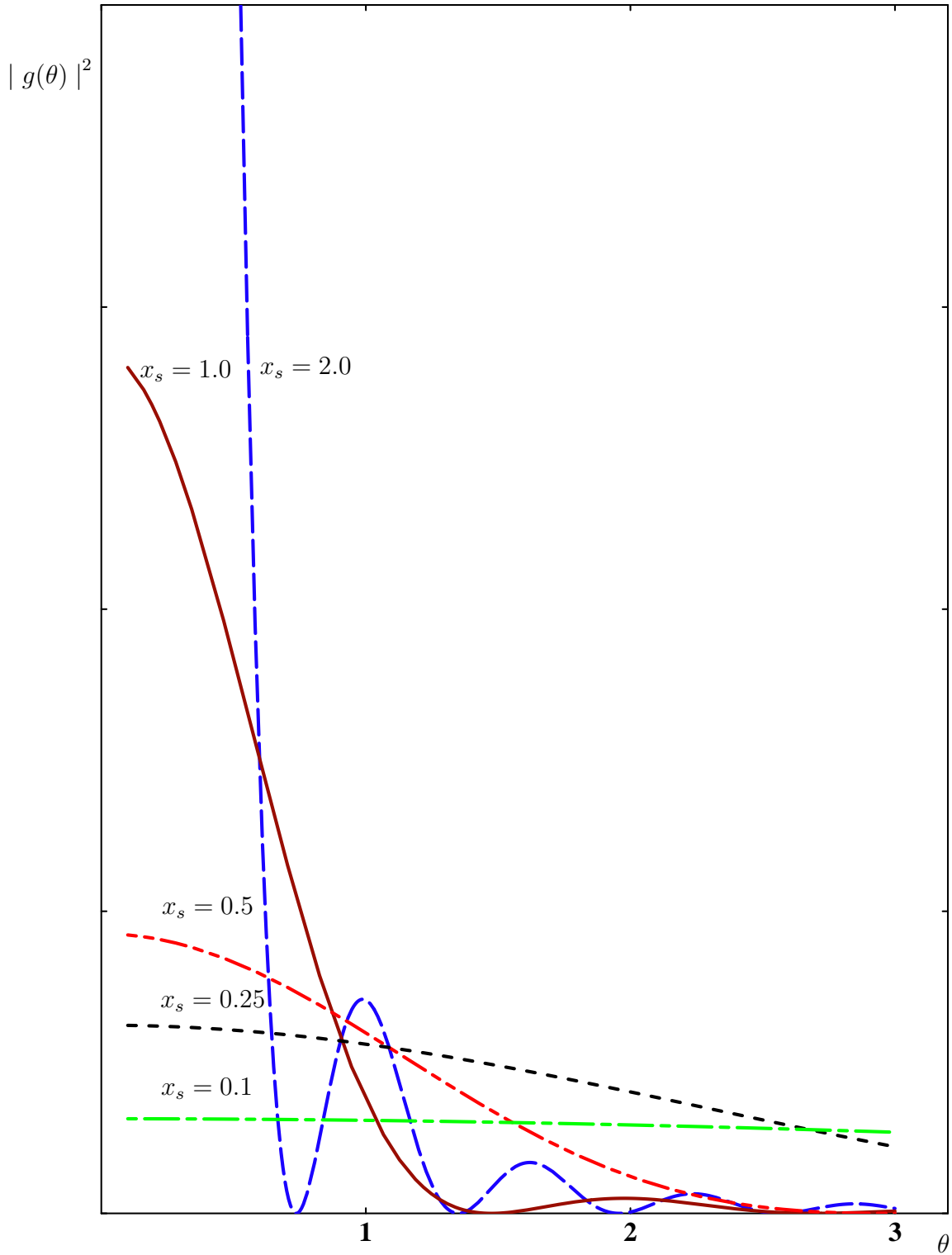}
\caption[fig4]{ANGULAR DISTRIBUTION $\mid g(\theta) \mid ^2 $ OF ABSORBED WAVES\label{fig4}}
\end{figure}

The angular distribution of absorbed waves is shown in [\ref{fig4}]. It is isotropic for low frequencies and gradually shows features of a diffraction pattern, as the frequency increases. It presents an absolute maximum in the forward direction which grows and narrows as the frequency increases. In the geometrical-optics limit, this results in a Dirac Delta  distribution. The analytic behaviour expresses in terms of the Bessel function $ {\mathcal{J}}_{1} $, as given by eq. (\ref{bessel}) below.

In the course of this research, we have developed accurate and useful computational methods based on the analytical resolution of the wave equation, which in addition, have allowed us to determine the range of validity of different approximations for low and high frequencies made by other authors (Starobinsky, Sov. Phys. JETP {\bf 37}, 1, 1973; Unruh, Phys. Rev. {\bf D14}, 3251, 1976), respectively, and by ourselves \cite{ns76}. It follows that the analytical computation of elastic scattering parameters for low frequencies is a rather open problem.

We have also obtained several properties concerning the scattering, absorption and emission parameters in a partial wave analysis. They are repported in references \cite{ns} and \cite{ns2}. Some of them are also reported in references \cite{ns79} and \cite{fu}. 

The work presented here has also a direct interest for the field and string quantization in curved space-times, related issues and other current problems. See the Conclusions Section at the end of this paper. 

\section{PARTIAL WAVE ANALYSIS}

{\margen} The partial scattering matrix is given by
\be
S_l = e^{2i {\delta}_l}
\ee
\be
{\delta}_l = {\eta}_l + i {\beta}_l
\ee

We have found (\cite{ns}, \cite{ns2}) that the real and imaginary parts of the Black Hole phase shifts $ \delta_l $ are {\underline{odd}} and {\underline{even}} functions of the frequency respectively:
\bes
\eta_l (x_s) = - \eta_l(-x_s) & & \\ 
\beta_l(x_s) = \beta_l(-x_s) & , & x_s \equiv k r_s = 2 k M \nonumber
\ees

\pagebreak[4]

In terms of the phase shifts, the partial elastic and absorption cross sections are respectively given by:
\bee
\epsilon_l = {{\pi}\over{{x_s}^2}} (2 l + 1) (1 - e^{-2 \beta_l}\cos{2 \eta_l} + e^{-4 \beta_l}) & & \mathrm{partial\; elastic \; cross \; section}\\
\sigma_l = {{\pi}\over{{x_s}^2}} (2 l + 1) (1 - e^{- 4 \beta_l}) \mathrm{\hspace{2.5cm}}  & & \mathrm{partial \; absorption \; cross \; section}
\eee
\subsection{Absorption cross sections}

{\margen} For all value of the angular momentum $ l $, the imaginary part of the phase shifts, $\beta_l(x_s)$, is a monotonically increasing function of $x_s$.

All $\beta_l(x_s)$ are zero at $x_s = 0$ and tend to infinity linearly with $x_s$ as $x_s$ increases to infinity.

\addcontentsline{toc}{subsubsection}{Low frequencies}
\noindent {\bf \underline{Low frequencies}}: For low frequencies, $(x_s \ll 1)$, $\beta_l(x_s)$ behaves as:
\bes
{\beta_l}^{exact}(x_s \ll 1) = C_l {x_s}^{2 l + 2} & \mathrm{\hspace{1.5cm}} &  x_s \ll 1 \; , \; x_s \equiv k \: r_s  \nonumber \\
C_l = {{2^{2 l} (l !)^6}\over{{[(2l)!]}^2 {[(2 l + 1)!]^2}}} & &
\ees

We have found for $C_l$ values in agreement with Starobinsky's formulae (Starobinsky, Sov. Phys. JETP {\bf 37} (1973) 1), for $x_s = 0$ and $l = 0$. However, the Starobinsky's approximation is accurate only in a small neighborhood of $x_s = 0$. For example, the ratio
\bee
{{{\beta_0}^{exact}(x_s) - C_0 {x_s}^2}\over{{\beta_0}^{exact}(x_s)}} \equiv X_0
\eee
varies as
\bee
0.15 \leq X_0 \leq 0.5 & \mathrm{\hspace{7mm} for \hspace{7mm}} & 0.05 \leq x_s \leq 0.1
\eee
For $ l = 1 $:
\bee
0.18 \leq X_1 \leq 0.6 & \mathrm{\hspace{7mm} for \hspace{7mm}} & 0.05 \leq x_s \leq 0.1
\eee
For small $x_s$, the inaccuracy of Starobinsky's approximation increases with $ l $.

At $x_s = 0$, all absorption cross sections $\sigma_l(x_s)$ are zero, except for $l=0$. For the S-wave:
\beq
\sigma_0(0) = 4 \pi
\eeq

The presence of a pole at $x_s = 0$ for $l \geq 1$ in the Jost function of the Black Hole (\cite{ns}, \cite{ns2}) means that waves with very small frequency and $ l \neq 0 $ are repelled out of the vecinity of the black hole.

\addcontentsline{toc}{subsubsection}{High frequencies}
\noindent {\bf \underline{High frequencies:}} For high frequencies, $(x_s \gg 1)$, the imaginary part of the phase shifts are given by 
\bes
{\beta_l}^{exact}(x_s) = {\beta_l}^{as}(x_s) + O({{l + {1/2}}\over{{x_s}^{3/2}}}) & \mathrm{\hspace{1cm}} & x_s \gg 1
\ees 
${\beta_l}^{as} $ is the asymptotic expression derived with the DWBA (Distorted Wave Born Approximation):
\bes
{\beta_l}^{as}(x_s) = {\pi x_s} - {1/4} \ln 2 - 1/{16}{({\pi}/{x_s})}^{1/2} - {{{\pi}\over{2 \sqrt{2}}}} {{{(l + {1/2})}^2}\over{x_s}}
\ees
There is very good agreement between ${\beta_l}^{exact}$ and ${\beta_l}^{as}$. For example:
\bee
{\beta_0}^{exact}(2) = 5.79 & , & {\beta_0}^{as}(2) = 5.89 \\
{\beta_1}^{exact}(2) = 4.98 & , & {\beta_1}^{as}(2) = 4.78
\eee
For all $x_s$, $ \beta_l(x_s) $ are described in ref. \cite{ns}.

The differential absorption cross section per unit solid angle $d \Omega$ for the Black Hole
\bee
{{d \sigma_A(\theta)}\over{d \Omega}} = {|\sum_{l=0}^{\infty} (2 l + 1) g_l(x_s) P_l(\cos \theta)|}^2 
\eee
is expressed in terms of the Bessel function $J_1$ as \cite{ns2}:
\beq
\label{bessel}
{{d \sigma_A(\theta)}\over{d \Omega}} \ \  \stackrel{x_s \gg 1 \; \theta \rightarrow 0}{\simeq} \ \  {{{27}\over{4}} {{{x_s}^2}\over{{\theta}^2}} {[J_1({{\sqrt{27}}\over{2}} x_s \theta)]}^2}
\eeq

\vspace{-2mm}
\subsection{Elastic Scattering}
\vspace{-2mm}
{\margen} For all angular momenta $l \neq 0$, the real part of the phase shifts, $\eta_l(x_s)$, as a function of $x_s$ has three zeros in the range $ 0 \leq x_s \leq \infty $ \cite{ns2}.

For the S-wave, $\eta_0(x_s)$ has two zeros.

We denote as ${x_s}^i(l)$, the frequencies at which $\eta_l(x_s)$ vanishes ($ i = 0, 1, 2 $ stands for first, second or third zero, respectively). 
\bee
{\eta_l}({x_s}^{l_(i)}) \ = \ 0 & {,} & \ \ i = 1, 2, 3. 
\eee 
The first zero of $\eta_l$ is at $x_s \; = \; 0$.
\pagebreak[3]

\addcontentsline{toc}{subsubsection}{Low frequencies}
\noindent {\bf \underline{Low frequencies:}} For low $x_s$, $x_s \ll 1$:
\bes
\eta_l(x_s) \   \stackrel{x_s \ll 1}{\simeq} \  a_l \: x_s & ,{\mathrm{\hspace{1cm}}} & {\mathrm{for \  all \  l}} 
\ees

The detailed behavior is discussed in ref.\cite{ns2}. Usually, the presence of long-range interactions produces a divergence in the low-frequency behaviour of the phase shifts. This is not the case here since the Coulomb interaction vanishes when the wave energy tends to zero.

In the interval $ 0 \leq x_s \leq {x_s}^{(1)}(l) $ :
\bee
\eta_l < 0 \ \ &  & \ \ {\mathrm{for \; low \;}} l \neq 0 \\
\eta_0 > 0 \ \ &  & \ \ {\mathrm{for\;}} l = 0
\eee
and very nearly equal to zero. 

For increasing $ l $, $ \eta_l $ becomes more and more negative according to the general variation $ \Delta \eta_l(x_s) $, due to variations $ \Delta V_{eff} $ of the effective potential:
\be
\Delta \eta_l = - {1\over{k}} \int_{0}^{\infty} {{dr}\over{1-{{r_s}\over{r}}}} {({{R_l}\over{r}})}^2 \Delta V_{eff}
\ee
where $R_l$ is the solution to radial wave equation and
\bee
V_{eff}(x^*)=(1-{{x_s}\over{x}})\left[{{x_s}\over{x^3}} + {{l(l+1)}\over{x^2}}\right] \ & , & x^* = x + x_s \ln(1-{{x_s}\over{x}}) \; \; .
\eee

\addcontentsline{toc}{subsubsection}{High frequencies}
\noindent {\bf \underline{High frequencies:}} For large $x_s$, $x_s \gg 1$:
\bes \label{dieza}
{\eta_l}^{exact}(x_s) = {\eta_l}^{as}(x_s) + O({({{1}\over{x_s}})}^{3/2}) \  & , & \ \  l \ll x_s
\ees
where
\bes
{\eta_l}^{as}(x_s) & = & {{\delta_c}\over{2}} - x_s + {{\pi}\over{2}}(l + {3\over{2}}) + {1\over{16}}{({{\pi}\over{x_s}})}^{1/2} + O({{1}\over{{x_s}^{3/2}}}) \label{diezb} \\ 
\delta_c & = & Im \ln \Gamma({1\over{2}} - 2 i\:  x_s) \nonumber \\
& = & - 2 x_s \ln({{2 x_s}\over{e}}) + O({{1}\over{x_s}}).\nonumber
\ees
${\eta_l}^{as}$ is the asymptotic formula derived by us in the DWBA (Distorted Wave Born Approximation) scheme \cite{ns76}.

\noindent ${\eta_l}^{as}$ and ${\eta_l}^{exact}$ are in good agreement. For example,
\bee
{\eta_0}^{exact}(2.3) = - 1.2 & , & {\eta_0}^{as}(2.3) = - 1.18 \\
{\eta_0}^{exact}(2.5) = - 1.6 & , & {\eta_0}^{as}(2.5) = - 1.60
\eee
We have
\bee
\eta_l(x_s \rightarrow \infty) \rightarrow - x_s \ln(2 x_s) \rightarrow - \infty & , & {\mathrm{for\; fixed \;}} l {\mathrm{\; and \;}} x_s \rightarrow \infty \\
\eta_l(x_s) \rightarrow - x_s \ln{(l + {1\over{2}})} \mathrm{\hspace{2cm}} & , & {\mathrm{for \; fixed \;}} x_s {\mathrm{\; and \;}} l \rightarrow \infty
\eee

\addcontentsline{toc}{subsubsection}{High Angular Momenta}
\noindent {\bf \underline{High Angular Momenta:}} High-order partial waves give an important contribution to the elastic scattering amplitude. This fact lead us to study $\eta_l(x_s)$ for $l \geq 2$ and fixed $x_s$. We found:
\bee
\eta_l(x_s) \; = \; {\eta_l}^{coul}(x_s) + \alpha_0(x_s) + \Delta_l(x_s) \ \ & , & \ \ l \; \gg  \; x_s
\eee
where
\bes
{\eta_l}^{coul}(x_s) & = & \arg{ \Gamma(l+1-ix_s)}  \\
\Delta_l(x_s) & = & {{\alpha_1(x_s)}\over{(l+{1\over{2}})}} + {{\alpha_2(x_s)}\over{{(l+{1\over{2}})}^2}} + {{\alpha_3(x_s)}\over{{(l+{1\over{2}})}^3}} + O({1\over{{(l+{1\over{2}})}^4}}) \nonumber \\
\alpha_l(x_s) & \sim & {(x_s)}^{l+1} \nonumber \\
\alpha_0(x_s) & \simeq & {1\over{5}}x_s \; \; , \; \; \; \;
\alpha_1(x_s) \; \; \simeq \; \; {3\over{2}}{x_s}^2 \nonumber 
\ees
As could be expected, the leading behaviour of $\eta_l(x_s)$ for large $ l $ is the same as that of the Coulomb phase shift $ {\eta_l}^{coul} $.

The difference
\bee
[{\eta_l}^{exact}(x_s) - {\eta_l}^{coul}(x_s)] = x_s f(b)
\eee
can be written as $ x_s $ times a function $ f(b) $ of the impact parameter $ b = {{(l+{1\over{2}})}\over{x_s}} $, which is analytic at $ b = 0 $.

The fact that $ \eta_l $ does not vanishes in the infinite energy limit is a consequence of the strong attractive character of the black hole interaction at short distances (the term of the type $ - {{1\over{{(r-r_s)}^2}} {({k_{r_s}}^2 + {{{r_s}^2}\over{4 r^2}})}} $ in the radial wave equation).

\section{DIFFERENTIAL ELASTIC CROSS SECTION}

{\margen} The scattering amplitude, whose squared modulus gives the differential elastic cross section, is given by
\bee
f(\theta) = \sum_{l=0}^{\infty} {{(2l+1)}\over{2ix_s}}(\exp{(-2\beta_l)}\exp{(2i\eta_l)}-1)P_l(\cos{\theta})
\eee

For all values of $x_s$ and small $\theta$, the differential elastic cross section behaves as
\beq
{\mid f(\theta) \mid}^2 = {{4}\over{{\theta}^4}} - {{C_1(x_s)}\over{{\theta}^3}} - {{4/3}\over{{\theta}^2}} + {{C_2(x_s)}\over{\theta}} + C_3(x_s) + O(\theta)
\eeq

The expressions for $ C_i(x_s), \; i = 1, 2 , 3 $ are given by \cite{ns2}:
\bee
C_1(x_s) & = & {{8 {\alpha}_1}\over{x_s}} \cos{(2 {\gamma}_{(-)})} \\
C_2(x_s) & = & \left (1 + {4\over{x_s}} - {{15}\over{4 {x_s}^2}}\right) 2 {\alpha}_1 \sin{(2 {\gamma}_{(-)})} + \\
& + & \left(1 + {1\over{4 {x_s}^2}}\right) {\alpha}_1 \sin{(2 {\gamma}_{(+)})} + \\
& - & 4 {{\alpha}_1}^3 \cos{(2 {\gamma}_{(-)})} \\
C_3(x_s) & = & {{{x_s}^2}\over{18}} + {4\over3}{{{\alpha}_1}\over{{x_s}^2}} + {{363}\over{72}} + {{97}\over{288}}{1\over{{x_s}^2}} + \\
& - & {7\over{6}}{{{\alpha}_2}\over{{x_s}^3}} + {{{{\alpha}_1}^4 - {{\alpha}_2}^2}\over{{x_s}^4}}
\eee
where
\bee
{\gamma}_{(\pm)} & = & \arg{\left(\Gamma({1\over{2}}+{i\:{x_s}}) \pm \Gamma({i\:{x_s}})\right )}
\eee
\bee
{\alpha}_0  \sim  {1\over{5}} x_s {\mathrm{\hspace{3mm}, \hspace{7mm}}} & 
{\alpha}_1  \sim  {3\over{2}} {x_s}^2 {\mathrm{\hspace{3mm}, \hspace{7mm}}} &
{\alpha}_2  \sim  {7\over{4}} {x_s}^3  
\eee

For intermediate angles, ${\mid f(\theta) \mid}^2$ has a complex behaviour with peaks and drops which disappear at the geometrical-optics limit.

\section{HAWKING EMISSION RATES}

{\margen} The energy emitted by the black hole in each mode of frequency $ k $ and angular momentum $ l $ is given by the Hawking's formula \cite{haw}
\bee
dH_l(k) = {{{\mathcal{P}}_l(k)}\over{(\exp{(4 \pi k r_s)} - 1)}} {{(2l + 1)}\over{\pi}}k\;dk
\eee

Let us recall our first analytic expression for the partial absorption rate $ {\mathcal{P}}_l(k)$ (S\'{a}nchez, 1976) \cite{ns76}. This is a formula for high frequencies $(k r_s \gg 1)$ obtained within the DWBA scheme:
\bes
{\mathcal{P}}_l(k) \; \; \stackrel{kr_s \gg 1}{=} \; \; {{1 - \exp{(-4 \pi k r_s)}}\over{1 + \exp{(- 4 \pi k r_s)}}} & & {\mathrm{for \;}} l \ll k r_s  \nonumber\\
\vspace{-6pt} \\
\vspace{-18pt}{\mathcal{P}}_l(k) \; \; \stackrel{kr_s \gg 1}{=} \; \; {1\over{1 + \exp{\{(2l+1)\pi[1-{{27k^2{r_s}^2}\over{{(2l+1)}^2}}]\}}}} &  & {\mathrm{for \;}} l \gg 1 \nonumber
\ees

In terms of this formula, $dH_l(k)$ can be expressed very simply by
\bes
dH_l(k) &  \stackrel{k r_s \gg 1}{=} & 
	{{1\over{(\exp{(4\pi k r_s)} + 1)}} {{2l +1}\over{2 \pi}} k dk} {\mathrm{\hspace{4.5cm}}}  l \ll k r_s \nonumber \\
\\
dH_l(k)	& \stackrel{k r_s \gg 1}{=} &
	 {{(2l+1) k dk}\over{(\exp{(4 \pi k r_s)} - 1)\{1+\exp{[(2l+1)\pi(1 - {{27k^2{r_s}^2}\over{{(2l+1)}^2}})]}\}}} {\mathrm{\hspace{5mm}}} l \gg 1 \nonumber
\ees
 
In order to compute the total emission $H(k)$, the total absorption $\sigma_A(k)$ eq.\ref{sigma}, and the set of properties discussed in the preceding section are needed.

By using the absorption cross section eq.\ref{sigma} and the partial wave results reported in the preceding sections, we have calculated the Hawking emission for a wide range of frequency and angular momentum (S\'anchez, 1978) \cite{ns}.

The total emission spectrum as a function of $x_s$ is plotted in Fig.\ref{fig1}. It does not show any of the interference oscillations characteristic of the total absorption cross section eq.\ref{sigma}, Fig.\ref{fig2}.\pagebreak[4] This is related to the fact that the S-wave contribution  predominates in Hawking radiation. For example, the maxima of $H_l(k)$ for $ l = 0, 1, 2 $ are in the ratio
\be
1 \ : \ {1\over{11}} \ : \ {1\over{453}}.
\ee

For angular momenta higher than two, $H_l(k)$ is extremely small.

The spectrum of total emission has only one peak following closely the S-wave absorption cross section behaviour. Its maximum lies at the same point the maximum of $\sigma_0$ $({x_s}^{max} = 0.23)$.

The peaks of $\sigma_1$ and $\sigma_2$ turn out to have no influence on $H(k)$.

In conclusion, Hawking emission is only important in the frequency range 
\beq
0 \ \leq \ k \  \leq {1\over{r_s}}
\eeq

\section{REMARKS ON APPROXIMATIONS}

\margen The analytic computation of the real part of the phase shift for low frequencies is a rather non-trivial problem.

Although the power-series expansion around $x_s = 0$ for the regular solution is known \cite{ocho}, the phase shifts cannot be obtained directly from it, because the asymptotic limit $r\rightarrow \infty$ cannot be taken term by term. Thus, the phase shifts are usually calculated by a standard procedure in which the radial equation is solved approximately in two or more regions of the positive real axis. By matching these solutions in the overlapping regions, approximated expressions for the phase shifts are found. By this procedure, several authors (Starobinsky \cite{nueve} and Unruh\cite{diez}) have obtained approximate expressions for the imaginary part of the phase shifts.

In Starobinsky's approximation, effects connected to the Coulomb tail of the interaction \underline{have not} been taken into account. With his approximation, it is possible to find for the real part of the phase shift a linear behaviour in $x_s$ $(\eta_l \sim a_l x_s)$, but inaccurate values for the coefficient $a_l$ are obtained.

In the context of a massive field, Unruh included the Coulomb interaction. However, his approximation is not sufficiently accurate to give the real part of the phase shift at least for $l=0$ in the zero-mass case.\pagebreak[3]

The discrepancy between Unruh's approximation with the exact calculation can be explained as follows: in Unruh's approach, for $r \gg r_s$, all terms of order higher than ${({{r_s}\over{r}})^2}$ are neglected in the exact radial wave equation written as
\beq \label{adob}
{{d^2}\over{dr^2}}(\xi R_l) + \left [ k^2 + {{2 k^2 r_s}\over{r-r_s}} + 
{{k^2{r_s}^2}\over{{(r-r_s)}^2}} + {1\over{4}} {{{r_s}^2}\over{r^2{(r-r_s)}^2}} - {{l(l+1)}\over{r{(r-r_s)}}} \right ] (\xi R_l) = 0
\eeq
where
\be
\xi = {1\over{k{[r(r-r_s)]}^{1\over{2}}}}.
\ee

In Unruh's approximation, the exact solutions (Coulomb functions) of the approximate wave equation for $r \gg r_s$ are also used in the region $kr \ll 1$. However, the term ${{{1\over{4}}{r_s}^2}\over{r^2{(r-r_s)}^2}}$ is much smaller than the Coulomb term, only for $k^2 r^2 \gg {1\over{8}}{{r_s}\over{r}}$. Then the following double inequality must hold:
\beq \label{doble}
1 \ \ \gg \ \ (kr) \ \  \gg \ \ {{{(kr_s)}^{1\over{3}}}\over{2}}
\eeq
\margen Thus, in his approximation, one cannot expect to obtain good results for $x_s$ which are not extremely small. [If one consideres that the symbol $\gg$ indicates a difference of one order of magnitude, inequality (\ref{doble}) implies $x_s \leq 10^{-6}$].

Finally, concerning the ${({{r_s}\over{r}})}^2$ order term ${{k^2\;{r_s}^2}\over{{(r-r_s)}^2}}$ in Eq.(\ref{adob}), it cannot be neglected obviously for $l=0$.

In a different approach (Persides, \cite{ocho}), it has been shown that the Wronskian of the exact radial wave equation can be expanded in a double power series of $x_s$ and $x_s \ln x_s$, although explicit expressions for the coefficients are not known. In this way, if we calculate the phase shifts the leading behaviour obtained for the real part $\eta_l$ at low frequencies shows a linear and a cubic term in $x_s$ plus $O({x_s}^3 \ln x_s)$.

For large values of $x_s$, ($x_s > {x_s}^{(2)}(l)$, where ${x_s}^{(2)}(l)$ is the frequency at which the second zero of $\eta_l$ occurs), it follows from our results \cite{ns2} that $\eta_l$ is negative and a monotonically decreasing function os $x_s$. Here ${({x_s}^{(2)})}^2 \gg V_{eff}(max)$ and $\eta_l$ tends to $- \infty$ as $x_s$ increases to $\infty$.

For large $x_s$, we find a good agreement between our exact values eq.(\ref{dieza}) and the asymptotic formula eq.(\ref{diezb}) derived by us \cite{ns76} in the Approximation DWB (Distorted Wave Born Approximation).  

\section{CONCLUSIONS}

\margen Accurate and powerful computational methods, based on the analytic
resolution of the wave equation in the black hole background, developed by
the present author allow to obtain the total absorption spectrum of the
Black Hole. As well as phase shifts and cross sections (elastic and
inelastic) for a wide range of energy and angular momentum, the angular
distribution of absorbed and scattered waves, and the Hawking emission
rates.

The total absorption spectrum of the Black Hole is known exactly.
The absorption spectrum as a function of the frequency shows a 
{\bf{remarkable oscillatory}} 
behaviour characteristic of a diffraction pattern. The absorption cross
section oscillates around its optical geometric limit with decreasing 
amplitude and almost constant period. Such oscillatory absorption pattern is 
an unique distinctive feature of the Black Hole. Absorption by ordinary bodies,
complex refraction index or optical models do not present these features.

For ordinary absorptive bodies, the absorption takes place in the whole medium
while for the Black Hole it takes place only at the origin ($ r = 0 $).

For the Black Hole, the effective Hamiltonian describing the wave-black hole
interaction is non-hermitian, despite of being real, due to its singularity
at the origin $(r = 0)$. The well known unitarity (optical)
theorem of the potential scattering theory can be generalized to the Black
Hole case, explicitly relating the presence of a non zero absorption cross
section to the existence of a singularity $(r=0)$ in the space time. 

All partial absorption amplitudes have absolute maxima at the frequence 
$k = {{3 \sqrt{3}}\over{4 M}} (l + {{1}\over{2}})$. By summing up all angular 
momenta, each absolute partial wave maximum, produces a relative maximum
in the total cross section giving rise to the presence of oscillations.

All these results allow to {\bf understand} and {\bf reproduce} the exact
absorption spectrum in terms of the Fresnel-Kirchoff diffraction theory.
The oscillatory behaviour of ${\sigma}_A(k)$ is due to the interference of the 
absorbed rays arriving at the origin  $(r = 0)$ through different optical
paths.

Semiclassical WKB Approximation for the Scattering by Black Holes only gives
information about the high frquency $(kr_s \gg 1)$, (and high angular momenta),
of the elastic (real part) of the phase shifts, but fail to describe well the 
absorption properties (and low partial wave angular momenta).

DWBA (Distorted Wave Born Approximation) for the Black Hole as it was
implemented by the present author more than twenty years ago \cite{ns76} is
an accurate better approximation for high frequencies $(k r_s \gg 1)$ to
compute the absorption (imaginary part) phase shifts and rates, both for
high $(l \gg k r_s)$ and low $(l \ll k r_s)$ angular momenta.

Approximative expressions (whatever they be), for very high frequencies, or for
low frequencies {\bf{do not}} allow to find the remarkable
{\bf{oscillatory}} behaviour of the total absorption cross section
as a function of frequency, of the Black Hole. 
The knowledge
of the highly non trivial total absorption spectrum of the Black Hole needed 
the development
of computational methods \cite{ns} more powerful and accurate than the commonly used approximations.

The angular distribution of absorbed and elastically scattered waves have been also computed with these methods.

The conceptual general features of the Black Hole Absorption spectrum will 
survive for higher dimensional ($D > 4$) generic Black Holes, and including
charge and angular momentum. They will be also present for Black Hole 
backgrounds solutions of the 
low energy effective field equations of string theories and D branes.

The Absorption Cross Section is a classical concept. It is exactly known and 
understood in terms of classical physics (classical perturbations around
fixed backgrounds). (Although, of course, it is possible to rederive and 
compute magnitudes from several different ways and techniques).

An increasing amount of paper \cite{once} has been devoted to the 
computation of absorption cross sections (``grey body factors'') of Black 
Holes, whatever D-dimensional, ordinary, D-braneous, 
stringy, extremal or non extremal. All these papers \cite{once} deal
with approximative expressions for the partial wave
cross sections. In all these papers \cite{once} the fundamental remarkable 
features of the Total Absorption Spectrum of the Black Hole are overlooked.


\begin{thebibliography}{9}

\bibitem{haw} S.Hawking, {\it ``Particle Creation by Black Holes''}, Comm. Math. Phys. {\bf 43} (1975) 199.
\bibitem{ns} N.S\'{a}nchez, {\it ``Absorption and Emission Spectra for a Schwarzschild Black Hole''}, Phys.Rev.{\bf D18} (1978) 1030.
\bibitem{ns77} N.S\'{a}nchez, {\it ``Wave Scattering Theory and the Absorption Problem for a Black Hole''}, Phys.Rev.{\bf D16} (1977) 937.
\bibitem{ns2} N.S\'{a}nchez, {\it ``Elastic Scattering of Waves by a Black Hole''}, Phys.Rev.{\bf D18} (1978) 1798.
\bibitem{ns76} N.S\'{a}nchez, {\it ``Scattering of Scalar Waves from a Schwarzschild Black Hole''}, J.Math.Phys. {\bf 17} (1976) 688.
%\bibitem{st} A.Starobinsky, Sov.Phys.JETP {\bf 37} (1973) 1.
\bibitem{ns79} N.S\'{a}nchez, {\it ``Sur la Physique des Champs et la G\'{e}om\'{e}trie de l'Espace-Temps''}, Th\`{e}se d'Etat, Paris (1979).
\bibitem{fu} J.A.H.Futterman, F.A.Handler and R.A.Matzner, {\it ``Scattering from Black Holes''}, Cambridge University Press, Cambridge, U.K. (1988), and references therein.
\bibitem{ocho} S.Persides, Int.Jour.Math.Phys.{\bf 48} (1976) 165; {\bf 50} (1976) 229.
\bibitem{nueve} A.A.Starobinsky, Zh.Eks.Teor.Fiz. {\bf 64} (1973) 48. (Sov.Phys.-JETP {\bf 37}(1973)1).
\bibitem{diez} W.G.Unruh, Phys.Rev.{\bf D14} (1976) 3251.
\bibitem{once} For example: \begin{description}
\item S.S.Gubser and I.R.Klebanov, Phys.Rev.Lett.{\bf 77} (1996) 4491; 
\item J.Maldacena and A.Strominger, Phys.Rev.{\bf D55} (1997) 861 and Phys.Rev.{\bf D56} (1997) 4975; 
\item S.S.Gubser, Phys.Rev.{\bf D56} (1997) 4984 and {\bf hep-th/9706100}; 
\item M.Civeti\v{c} and F.Larsen, Phys.Rev.{\bf D56} (1997) 4994 and {\bf hep-th/9705192}; 
\item I.R.Klebanov and S.D.Mathur, {\bf hep-th/9701187}; 
\item S.Das, A.Dasgupta and T.Sarkar, Phys.Rev.{\bf D55} (1997) 12; 
\item S.P. de Alwis and K.Sato, Phys.Rev.{\bf D55} (1997) 6181; 
\item R.Emparan, {\bf hep-th/9704204}; 
\item H.W.Lee, Y.S.Myung and J.Y.Kim, {\bf hep-th/9708099}.
\end{description}

\end{thebibliography}
\end{document}